\documentclass[aps,pra,twocolumn,amsmath,amssymb]{revtex4-2}
\usepackage{graphicx}
\usepackage{bm}
\usepackage{multirow}
\usepackage{ulem}
\usepackage{color}
\usepackage{hyperref}
\usepackage{cleveref}
\usepackage{comment}
\usepackage{braket}
\usepackage{adjustbox}
\usepackage{float}
\usepackage{subfig}
\usepackage{lipsum}
\usepackage{tikz}
\usetikzlibrary{arrows.meta, positioning, calc}
\usepackage{subcaption} 
\usepackage{quantumoptics}
\begin{document}
\title{A Standard Quantum Mechanical Treatment to Rationalize the Delayed-Choice Quantum Eraser}

\author{Vipul Badhan}
\email{vipulbadhan269@gmail.com}
\author{Pranvi$^{\$}$}
\author{Samreet Dhillon$^{\$}$}
\author{Bindiya Arora}
\email{bindiya.phy@gndu.ac.in}
\affiliation{Department of Physics, Guru Nanak Dev University, Amritsar, Punjab 143005, India}
\affiliation{$^{\$}$These authors contributed equally to this work.}

\begin{abstract}
Ever since the proposal of delayed choice quantum erasure and subsequent realization in the experiment by Kim et al. \cite{kim2000}, the interpretation and implications of delayed-choice experiments have remained a subject of intense foundational debate.
This paper resolves the apparent paradox attached to the experiment using standard quantum mechanics. Using an extended Mach-Zehnder interferometer which captures every operational feature of the original experiment, we show that choosing between which-path and erasure detectors is simply a choice of measurement bases, which does not rewrite a photon's past. 
Furthermore, by mapping the experiment to a two-way Stern-Gerlach framework, we prove that quantum erasure is an expected result of measuring entangled states, not a physical anomaly. 
Ultimately, through a pedagogical game, we illustrate that the illusion of retrocausality arises from asking illegitimate questions, and that a forward-in-time description is entirely sufficient to explain the logic.
\end{abstract}

\date{\today}
\maketitle

\section{Introduction}
More than forty years have passed since Wheeler introduced the delayed-choice quantum eraser \cite{wheeler1978}, and over twenty since Kim et al. first implemented it in the laboratory \cite{kim2000}. Nevertheless, the proper interpretation of the delayed-choice experiment remains heavily debated till date~\cite{qureshi2020,qureshi2021delayedchoice, chiou2023delayedchoice}. Opinions in the literature diverge sharply, ranging from descriptions of the effect as a challenge to standard notions of  space and time~\cite{chaves2018,sabine2021} to characterizations of it as an experimental paradigm~\cite{exp1,exp2} that has generated widespread conceptual confusion.  A substantial body of theoretical work has attempted to demystify these results by framing them within established quantum frameworks, such as EPR correlations, causal modeling, and basis transformations \cite{kastner2019delayed, aharonov2005, hiley2006, ellerman2015, fankhauser2017, englert1999, mohrhoff1999,terno2011,masi}.

The conceptual lineage of delayed-choice \textit{gedanken} experiments can be traced back to Heisenberg’s 1927 microscope thought experiment \cite{heisenberg1927}, which analyzed the fundamental limits of measuring an electron's position. In 1931, Weizsäcker provided a rigorous account of this setup\cite{weizsacker1931}, noting that a photon scattering off an electron conveys different information depending on the observer's configuration. Einstein \cite{einstein1931} and Hermann \cite{hermann1935} separately refined this approach suggesting that the choice of the measurement configuration can be delayed until after the physical interaction between the photon and electron had occurred. This insight established the framework for modern delayed-choice protocols. Wheeler revived and formalized this concept in 1978, introducing a family of delayed-choice arrangements, most notably the interferometer variant based on a Mach–Zehnder configuration \cite{wheeler1978}. The paradigm was further extended by Scully and Drühl in 1982 with the introduction of the quantum eraser \cite{scully1982}. They demonstrated that the loss of interference in a two-path setup is not caused by a invasive disturbance during measurement, but rather by the mere availability of `which-path' information. If the path taken by a system can be known, interference vanishes. Yet, if this path information is subsequently erased, even after the primary detection events, interference can be recovered via correlated joint measurements. Much like Wheeler's original delayed-choice experiment, the delayed-choice quantum eraser has been the subject of extensive interpretational debate in both the physics literature and philosophical circles. 

A primary source of confusion in delayed-choice protocols stems from language that attributes intrinsic `which-path' or `both-paths' information to quantum states. Kastner forcibly argued \cite{kastner2019delayed}, the delayed-choice quantum eraser ``neither erases nor delays", pointing out that representing an entangled state vector in a specific basis does not imply that a particle possesses intrinsic information prior to measurement. Drawing a direct analogy to spin-$1/2$ EPR correlations, an electron detected along a particular measurement direction carries only outcome information along that axis. It is conceptually unsound to argue that a downstream measurement performed on a particle (idler) erases which-path information about its entangled counterpart (signal), given that the latter never possessed any  which-path information to begin with \cite{kastner2019delayed}. In a similar spirit, Gaasbeek emphasized the crucial distinction between correlation and causation \cite{gaasbeek2010demystifying}. Spacelike or delayed joint measurements exhibit backward correlations dictated by the initial state preparation, but these correlations strictly preclude backward causation. The joint probabilities are entirely symmetric under conditionalization ($\text{Prob}(A|B) \propto \text{Prob}(B|A)$), making the chronological sequence of registrations physically irrelevant to the observed statistics .

Other authors have sought to formalize these insights using alternative conceptual approaches. Chaves et al. \cite{chaves2018} applied causal modeling to modified delayed-choice arrangements, demonstrating that forward-in-time causal DAGs (directed acyclic graphs) fully account for the observed correlations without requiring temporal non-locality. Ionicioiu and Terno \cite{terno2011} replaced the classical choice of inserting or removing a beam splitter with a quantum ancilla in a superposition state. Adopting the operational definition of wave/particle duality as the ability/inability to produce interference~\cite{Greenstein1997} , they showed that wave and particle behaviors can be observed continuously within a single setup.

Qureshi analyzed the quantum eraser across channel configurations  \cite{qureshi2025enigma}, Stern–Gerlach models \cite{Qureshi2012}, and modified Mach–Zehnder configuration \cite{qureshi2021delayedchoice}. Qureshi argued that the delayed choice actually ``leaves no choice" for the experimenter: every registered signal photon fixes the state of the idler in a unique mutually unbiased basis. In this view, the role of mutually unbiased bases for which-path detectors has been historically overlooked, and path information is erased by the state projection itself \cite{qureshi2021delayedchoice}.  Chiou \cite{chiou2023delayedchoice} demonstrated that a modified Mach–Zehnder interferometer with asymmetric beam splitters shares the exact mathematical structure of an EPR-Bohm spin experiment. Chiou noted that claims of erasure rely on the unverified counterfactual presupposition that a signal photon must traverse a single localized path upon entering the apparatus, a presupposition that is merely one interpretation among many \cite{chiou2023delayedchoice}.

Although numerous setups have been proposed to clarify delayed-choice quantum erasure, most modify the underlying protocol. They often treat which-path detection and erasure as separate, two-step processes \cite{chiou2023delayedchoice,qureshi2021delayedchoice}, rely on alternative physical mechanisms \cite{scully2002}, or introduce external controlling ancillas \cite{terno2011} departing from the landmark experiment by Kim et al. \cite{kim2000}. In contrast, our proposed set-up inherently produces superposition between eraser (ability to produce interference) and which-path (inability to produce interference) behaviors in a single set-up without any external controlling ancillas, modeling the Kim et al. \cite{kim2000} experiment systematically. 
By avoiding retrocausal frameworks and heavy theoretical machinery, we demonstrate that these delayed-choice experiments and their apparent paradoxes can be fully resolved through straightforward, conceptually transparent mechanisms if one understands the role of mutually unbiased basis sets in such experiments. To ground this perspective, we present explanations using three distinct models:

\begin{enumerate}
    \item An extended, operational variation of the Mach–Zehnder interferometer.
    \item     An analogical treatment using two-way Stern–Gerlach measurements.
    \item A pedagogical game that replicates the delayed-choice outcomes.
\end{enumerate}

By explicitly defining the $50:50$ beam splitters and maintaining full symmetry across detectors $D_1$ through $D_6$, our model captures every operational feature of the Kim et al. experiment, including the simultaneous generation of interference  and anti-interference sub-ensembles. 
Furthermore, by providing an exact mathematical translation to a Stern–Gerlach spin-$1/2$ setup, we demonstrate that this erasure mechanism is not an anomaly, but a universal property of quantum measurement in mutually unbiased bases. Through the pedagogical game, we demonstrate that the apparent retrocausality is merely an illusion arising from ill-posed questions, and that a fully consistent forward-in-time description suffices to account for all observed behavior.


\section*{An extended Mach-Zehnder Interferometer setup}

\begin{figure}[htbp]
    \centering
    \resizebox{0.2\linewidth}{!}{\begin{tikzpicture}


\node[text=black, font=\large\bfseries] at (-0.75, 0.25) {$a$};
\node[text=black, font=\large\bfseries] at (0.25, -0.75) {$b$};

\node[text=black, font=\large\bfseries] at (0.75, -0.35) {$\gamma_1$};
\node[text=black, font=\large\bfseries] at (-0.35, 0.75) {$\gamma_0$};

\drawbeam{-1}{0}{0}{0}{1.5pt}    
\drawbeam{ 0}{-1}{0}{0}{1.5pt}    
\drawbeam{0}{0}{1}{0}{1.5pt}       
\drawbeam{0}{0}{0}{1}{1.5pt}       

\drawbeamsplitter{0}{0}{0}{1}{125,185,222}
\node[below left, text=black!80, inner sep=8pt] at (-0.1,-0.1) {BS};

\end{tikzpicture}}
    \caption{Schematic representation of a standard beam splitter $BS$. Input spatial modes $\ket{a}$ and $\ket{b}$ undergo a unitary transformation, yielding output spatial modes $\ket{\gamma_0}$ and $\ket{\gamma_1}$}. 
    \label{fig:bs}
\end{figure}
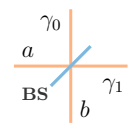
\textit{\emph{Action of a beam splitter}:}
Before examining the delayed-choice architecture, it is instructive to define the unitary transformation executed by a standard $50:50$ non-polarizing beam splitter ($\text{BS}$). 
A beam splitter divides the incident wave function into two distinct spatial modes, thereby generating a superposition state. For a lossless 50:50 beam splitter, the input states evolve according to the following transformations:
\[
|a\rangle \rightarrow \frac{1}{\sqrt{2}} |\gamma_1\rangle + i\frac{1}{\sqrt{2}} |\gamma_0\rangle,
\]
\[
|b\rangle \rightarrow i\frac{1}{\sqrt{2}}|\gamma_1\rangle + \frac{1}{\sqrt{2}}|\gamma_0\rangle.
\]
The phase factor $i$ in this transformation is essential for maintaining the unitarity of the beam-splitter matrix, which guarantees that orthogonal input states remain orthogonal after the transformation.
 This coherent splitting of the single-photon wave function across two spatially distinct trajectories forms the foundational building block for path-based interferometry.

\textit{\emph{Detector states and measurement notation}:}
To explicitly account for the measurement apparatus, an overall detector state $|D\rangle$ is included at each step. This vector represents the ``off" composite state of all six detectors used in our set-up:
\begin{equation}
    |D\rangle = |D_1\rangle |D_2\rangle |D_3\rangle |D_4\rangle |D_5\rangle |D_6\rangle.
\end{equation}
A detection event (or ``click'') at the $n$-th detector is represented by the transformation:
\begin{equation}
     |d_n\rangle |D_n\rangle\rightarrow|\overline{D_n}\rangle ,
\end{equation}
where $|d_n\rangle$ denotes the photon state in the $d_n$ path, and $|D_n\rangle$ represents the individual detector's off state.

\textit{\emph{Standard Mach-Zehnder setup}:}

\begin{figure}[htbp]
    \centering
    \resizebox{0.6\linewidth}{!}{\begin{tikzpicture}

\node[text=black, font=\large\bfseries] at (-0.75, 0.25) {$a$};
\node[text=black, font=\large\bfseries] at (0.25, -0.75) {$b$};

\node[text=black, font=\large\bfseries] at (2, -0.4) {$\gamma_1$};
\node[text=black, font=\large\bfseries] at (-0.4, 1.5) {$\gamma_0$};

\node[text=black, font=\large\bfseries] at (2, 3.3) {$\gamma_0$};
\node[text=black, font=\large\bfseries] at (4.3, 1.5) {$\gamma_1$};

\node[text=black, font=\large\bfseries] at (3.75, 3.4) {$d_1$};
\node[text=black, font=\large\bfseries] at (4.45, 2.75) {$d_2$};

\drawbeam{-1}{0}{0}{0}{1.5pt}    
\drawbeam{ 0}{-1}{0}{0}{1.5pt}    
\drawbeam{0}{0}{4}{0}{1.5pt}       
\drawbeam{0}{0}{0}{3}{1.5pt}       
\drawbeam{0}{3}{5}{3}{1.5pt}       
\drawbeam{4}{0}{4}{4}{1.5pt}     

\drawbeamsplitter{0}{0}{0}{1}{125,185,222}
\node[below left, text=black!80, inner sep=8pt] at (-0.1,-0.1) {BS1};

\drawbeamsplitter{4}{3}{0}{1}{125,185,222}
\node[below left, text=black!80, inner sep=8pt] at (3.9,2.9) {BS2};

\drawmirror{0}{3}{45}{1}{125,185,222}
\node[above left, text=black!80, inner sep=6pt] at (-0.1,3.1) {M1};

\drawmirror{4}{0}{225}{1}{125,185,222}
\node[below right, text=black!80, inner sep=6pt] at (4.1,-0.1) {M2};

\drawdetector{4}{4}{0}{1}{255,196,163}
\node[right, text=black!80, inner sep=8pt] at (4.2,4.2) {D1};

\drawdetector{5}{3}{-90}{1}{255,196,163}
\node[right, text=black!80, inner sep=8pt] at (5.2,3) {D2};

\drawphase{3}{3}{0}{1}{220,220,220}
\node[above, text=black!80, inner sep=12pt] at (3,3) {$\phi$};

\end{tikzpicture}}
    \caption{Standard Mach-Zehnder interferometer (MZI). An incident photon entering via input mode $\ket{a}$ is split by $\text{BS}_1$ into spatial modes $\ket{\gamma_0}$ and $\ket{\gamma_1}$. After reflection at mirrors $M_1$ and $M_2$ and undergoing a relative phase shift $\phi$, the paths recombine at $\text{BS}_2$. Output modes $\ket{d_1}$ and $\ket{d_2}$ are registered by detectors $D_1$ and $D_2$, respectively.}
    \label{fig:mzi}
\end{figure}
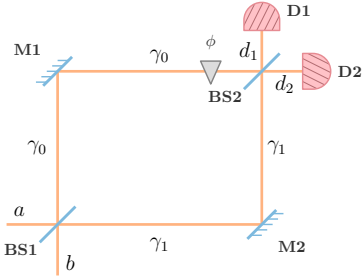
When two such beam splitters are aligned sequentially with two total-reflection mirrors, they form a standard Mach–Zehnder interferometer (MZI) {with phase difference $\Delta\phi$ between two arms}, as illustrated in Figure~\ref{fig:mzi}. An incident photon entering via input mode $\ket{a}$ is divided by $\text{BS}_1$ into a spatial superposition along the arms, $\ket{\gamma_0}$ and $\ket{\gamma_1}$. After acquiring an optical path phase difference $\Delta \phi$, the paths recombine at $\text{BS}_2$. The final state is given by:
\begin{equation}
\ket{\psi_{\text{MZI}}} = \frac{i}{2}\left(1 - e^{i\Delta \phi}\right)\ket{ \overline{D_1}} - \frac{1}{2}\left(1 + e^{i\Delta \phi}\right)\ket{ \overline{D_2}}
\end{equation}

For a balanced MZI  with ($\Delta \phi = 0$), destructive interference occurs at detector $D_1$, while constructive interference occurs at $D_2$, causing $D_2$ to register every single event. We designate this outcome as  \textit{interference}. Conversely, if a relative phase shift of $\Delta \phi = \pi$ is introduced, the interference condition inverts: constructive interference directs all photons exclusively to $D_1$ , leaving $D_2$ dark. We refer to this inverted signal as the \textit{complimentary-interference}, highlighting that photon detections shift to the detector that was completely dark in the balanced setup.

\textit{\emph{Extended Mach-Zehnder setup}:}
To demonstrate delayed-choice framework, we introduce an extended Mach–Zehnder architecture presented as case I (path-erasure), case II (which-way) and case III (delayed choice) as detailed in Figures~\ref{fig:interference}, \ref{fig:whichway},  and \ref{fig:delayedchoice}, respectively. An incident photon enters the primary beam splitter $\text{BS}_1$, creating a spatial superposition between paths $\gamma_0$ and $\gamma_1$. Entangled photon-pair source (EPPS) represented by the right-angled triangles, are positioned along these respective arms. When a photon traverses $\gamma_0$/$\gamma_1$, the EPPS generates a spatially entangled pair comprising a signal photon in arms $s_0$/$s_1$ and an idler photon in arms $i_0/i_1$. 

Immediately following  the action of EEPS, the global state of the system is described by the entangled state:
\begin{equation}
\ket{\psi_{\rm ES}} = \frac{1}{\sqrt{2}}\left( \ket{i_0}\ket{s_0} + \ket{i_1}\ket{s_1} \right)\ket{D}
\label{eq:entangled_initial}
\end{equation}
The signal paths ($s_0, s_1$) are routed toward the beam splitter $\text{BS}_2$, whose output ports lead directly to detectors $D_1$ (mode $d_1$) and $D_2$ (mode $d_2$); we designate this configuration as the ``signal MZI" for the remainder of the discussion. Meanwhile, mirrors $M_1$ and $M_2$ direct the corresponding idler paths ($i_0, i_1$) along longer optical routes, which is hereinafter referred to as the ``idler MZI".

\textit{\emph{Operational assumptions of the apparatus}:}
It is important to emphasize that we assume the BS and EPPS operate as lossless devices and they do not induce wave-function collapse or execute a path measurement and that polarization is entirely omitted from our treatment. Also, note that while positioning two independent EPPS in separate interferometric arms is unconventional in standard optical setups, it serves as a transparent proof-of-concept for generating non-separable path entanglement. The specific internal mechanism of pair creation, whether implemented via down-converting crystals~\cite{burnham1970}, atomic cascades~\cite{scully1982}, or polarization-to-path state conversion~\cite{kwiat1995}, is irrelevant to the conceptual conclusions of this work. 

\subsection{Case I}
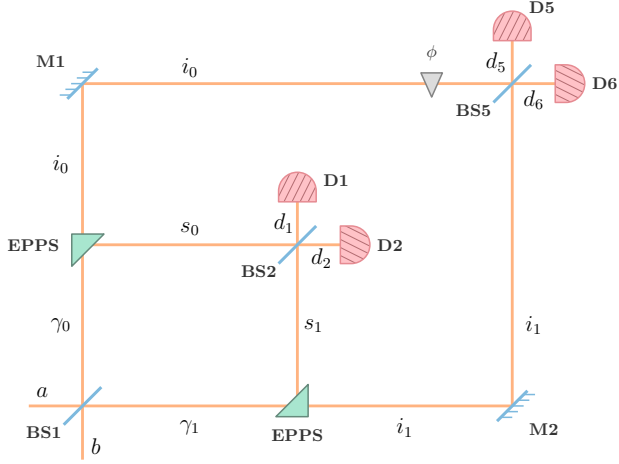
\begin{figure}[htbp]
    \centering
    \resizebox{\linewidth}{!}{\begin{tikzpicture}

\node[text=black, font=\large\bfseries] at (-0.75, 0.25) {$a$};
\node[text=black, font=\large\bfseries] at (0.25, -0.75) {$b$};

\node[text=black, font=\large\bfseries] at (2, -0.4) {$\gamma_1$};
\node[text=black, font=\large\bfseries] at (-0.4, 1.5) {$\gamma_0$};

\node[text=black, font=\large\bfseries] at (-0.4, 4.5) {$i_0$};
\node[text=black, font=\large\bfseries] at (2, 6.3) {$i_0$};

\node[text=black, font=\large\bfseries] at (8.4, 1.5) {$i_1$};
\node[text=black, font=\large\bfseries] at (6, -0.4) {$i_1$};

\node[text=black, font=\large\bfseries] at (2, 3.3) {$s_0$};
\node[text=black, font=\large\bfseries] at (4.3, 1.5) {$s_1$};

\node[text=black, font=\large\bfseries] at (3.75, 3.4) {$d_1$};
\node[text=black, font=\large\bfseries] at (4.45, 2.75) {$d_2$};

\node[text=black, font=\large\bfseries] at (7.7, 6.4) {$d_5$};
\node[text=black, font=\large\bfseries] at (8.4, 5.7) {$d_6$};

\drawbeam{-1}{0}{0}{0}{1.5pt}    
\drawbeam{ 0}{-1}{0}{0}{1.5pt}    
\drawbeam{0}{0}{8}{0}{1.5pt}       
\drawbeam{0}{0}{0}{6}{1.5pt}       
\drawbeam{0}{3}{5}{3}{1.5pt}       
\drawbeam{4}{0}{4}{4}{1.5pt}     
\drawbeam{0}{6}{9}{6}{1.5pt}      
\drawbeam{8}{0}{8}{7}{1.5pt}     

\drawbeamsplitter{0}{0}{0}{1}{125,185,222}
\node[below left, text=black!80, inner sep=8pt] at (-0.1,-0.1) {BS1};

\drawbeamsplitter{4}{3}{0}{1}{125,185,222}
\node[below left, text=black!80, inner sep=8pt] at (3.9,2.9) {BS2};



\drawbeamsplitter{8}{6}{0}{1}{125,185,222}
\node[below left, text=black!80, inner sep=8pt] at (7.9,5.9) {BS5};

\drawmirror{0}{6}{45}{1}{125,185,222}
\node[above left, text=black!80, inner sep=6pt] at (-0.1,6.1) {M1};

\drawmirror{8}{0}{225}{1}{125,185,222}
\node[below right, text=black!80, inner sep=6pt] at (8.1,-0.1) {M2};

\drawbbo{0.1}{2.9}{270}{1.2}{168,230,207}
\node[left, text=black!80, inner sep=12pt] at (0,3) {EPPS};

\drawbbo{3.9}{0.1}{90}{1.2}{168,230,207}
\node[below, text=black!80, inner sep=12pt] at (4,0) {EPPS};

\drawphase{6.5}{6}{0}{1}{220,220,220}
\node[above, text=black!80, inner sep=12pt] at (6.5,6) {$\phi$};

\drawdetector{4}{4}{0}{1}{255,196,163}
\node[right, text=black!80, inner sep=8pt] at (4.2,4.2) {D1};

\drawdetector{5}{3}{-90}{1}{255,196,163}
\node[right, text=black!80, inner sep=8pt] at (5.2,3) {D2};



\drawdetector{8}{7}{0}{1}{255,196,163}
\node[above right, text=black!80, inner sep=6pt] at (8.1,7.1) {D5};

\drawdetector{9}{6}{-90}{1}{255,196,163}
\node[right, text=black!80, inner sep=8pt] at (9.2,6) {D6};

\end{tikzpicture}}
    \caption{Extended MZI configuration with entangled photon pair sources (EPPS) situated along modes $\ket{\gamma_0}$ and $\ket{\gamma_1}$ generate signal ($\ket{s_0}, \ket{s_1}$) and idler ($\ket{i_0}, \ket{i_1}$) modes. Signal modes recombine at $\text{BS}_2$ leading to detectors $D_1$ and $D_2$. Idler modes undergo recombination at $\text{BS}_5$ yielding output modes $\ket{d_5}$ and $\ket{d_6}$ registered by detectors $D_5$ and $D_6$. Coincidence detections between $(D_1, D_5)$ and $(D_2, D_6)$ highlight complementary \textit{anti-interference} and \textit{interference} sub-ensembles.}
    \label{fig:interference}
\end{figure}

We first consider the setup in Figure~\ref{fig:interference}, where the idler mode $i_0$ is reflected by mirror $M_1$ and passes through a phase shifter setting $\phi = \pi$, while mode $i_1$ is routed directly via mirror $M_2$. Both idler modes then recombine at beam splitter $\text{BS}_5$, with output ports leading to detectors $D_5$ (mode $d_5$) and $D_6$ (mode $d_6$).  The phase shifter $\phi = \pi$ is introduced along mode $i_0$ for visual symmetry. Without this shift, constructive coincidence would pair $D_1$ with $D_6$ and $D_2$ with $D_5$. Introducing it, however, reverses the destructive phase conditions such that a detection at $D_1$ correlates neatly with $D_5$, and $D_2$ with $D_6$, providing a more intuitive pair mapping.

Applying all the transformations of the signal and idler MZI, the final state evaluates to (see Appendix I):
\begin{equation}
\ket{\psi_{\text{I}}} = \frac{i}{\sqrt{2}} \left( \ket{\overline{D_1}}\ket{\overline{D_5}} - \ket{\overline{D_2}}\ket{\overline{D_6}} \right)
\label{eq:psi_final_fig3}
\end{equation}
Equation~\eqref{eq:psi_final_fig3} reveals deterministic correlations. The coincidence count probability with respect to $D_6$ reveals that the signal MZI exhibits \textit{interference} i.e. only $D_2$ registers clicks, while the detection probability at $D_1$ is zero. Conversely, when considering the coincidence count probability with respect to $D_5$, the signal MZI displays \textit{complementary interference}, resulting in clicks exclusively at $D_1$. This demonstrates the wave-like nature of the photon.
\subsection{Case II}
\begin{figure}[htbp]
        \centering
        \resizebox{\linewidth}{!}{\begin{tikzpicture}

\node[text=black, font=\large\bfseries] at (-0.75, 0.25) {$a$};
\node[text=black, font=\large\bfseries] at (0.25, -0.75) {$b$};

\node[text=black, font=\large\bfseries] at (2, -0.4) {$\gamma_1$};
\node[text=black, font=\large\bfseries] at (-0.4, 1.5) {$\gamma_0$};

\node[text=black, font=\large\bfseries] at (-0.4, 4.5) {$i_0$};
\node[text=black, font=\large\bfseries] at (2, 6.3) {$i_0$};

\node[text=black, font=\large\bfseries] at (8.4, 1.5) {$i_1$};
\node[text=black, font=\large\bfseries] at (6, -0.4) {$i_1$};

\node[text=black, font=\large\bfseries] at (2, 3.3) {$s_0$};
\node[text=black, font=\large\bfseries] at (4.3, 1.5) {$s_1$};

\node[text=black, font=\large\bfseries] at (3.75, 3.4) {$d_1$};
\node[text=black, font=\large\bfseries] at (4.45, 2.75) {$d_2$};

\node[text=black, font=\large\bfseries] at (3.7, 6.4) {$d_3$};
\node[text=black, font=\large\bfseries] at (8.4, 2.7) {$d_4$};


\drawbeam{-1}{0}{0}{0}{1.5pt}    
\drawbeam{ 0}{-1}{0}{0}{1.5pt}    
\drawbeam{0}{0}{8}{0}{1.5pt}       
\drawbeam{0}{0}{0}{6}{1.5pt}       
\drawbeam{0}{3}{5}{3}{1.5pt}       
\drawbeam{4}{0}{4}{4}{1.5pt}     
\drawbeam{0}{6}{4}{6}{1.5pt}      
\drawbeam{8}{0}{8}{3}{1.5pt}     
\drawbeam{8}{3}{9}{3}{1.5pt}      
\drawbeam{4}{6}{4}{7}{1.5pt}     

\drawbeamsplitter{0}{0}{0}{1}{125,185,222}
\node[below left, text=black!80, inner sep=8pt] at (-0.1,-0.1) {BS1};

\drawbeamsplitter{4}{3}{0}{1}{125,185,222}
\node[below left, text=black!80, inner sep=8pt] at (3.9,2.9) {BS2};




\drawmirror{0}{6}{45}{1}{125,185,222}
\node[above left, text=black!80, inner sep=6pt] at (-0.1,6.1) {M1};

\drawmirror{8}{0}{225}{1}{125,185,222}
\node[below right, text=black!80, inner sep=6pt] at (8.1,-0.1) {M2};

\drawmirror{8}{3}{45}{1}{125,185,222}
\node[above left, text=black!80, inner sep=6pt] at (8,3) {M4};

\drawmirror{4}{6}{225}{1}{125,185,222}
\node[below right, text=black!80, inner sep=6pt] at (4,6) {M3};

\drawbbo{0.1}{2.9}{270}{1.2}{168,230,207}
\node[left, text=black!80, inner sep=12pt] at (0,3) {EPPS};

\drawbbo{3.9}{0.1}{90}{1.2}{168,230,207}
\node[below, text=black!80, inner sep=12pt] at (4,0) {EPPS};

\drawdetector{4}{4}{0}{1}{255,196,163}
\node[right, text=black!80, inner sep=8pt] at (4.2,4.2) {D1};

\drawdetector{5}{3}{-90}{1}{255,196,163}
\node[right, text=black!80, inner sep=8pt] at (5.2,3) {D2};

\drawdetector{4}{7}{0}{1}{255,196,163}
\node[above right, text=black!80, inner sep=6pt] at (4.1,7.1) {D3};

\drawdetector{9}{3}{-90}{1}{255,196,163}
\node[right, text=black!80, inner sep=8pt] at (9.2,3) {D4};



\end{tikzpicture}}
    \caption{Extended MZI configuration with idler modes $\ket{i_0}$ and $\ket{i_1}$ directed by mirrors $M_3$ and $M_4$ straight to detectors $D_3$ (mode $\ket{d_3}$) and $D_4$ (mode $\ket{d_4}$). A click at $D_3$ or $D_4$ provides  path information for the corresponding signal photon at $D_1$ or $D_2$.}
    \label{fig:whichway}
\end{figure}
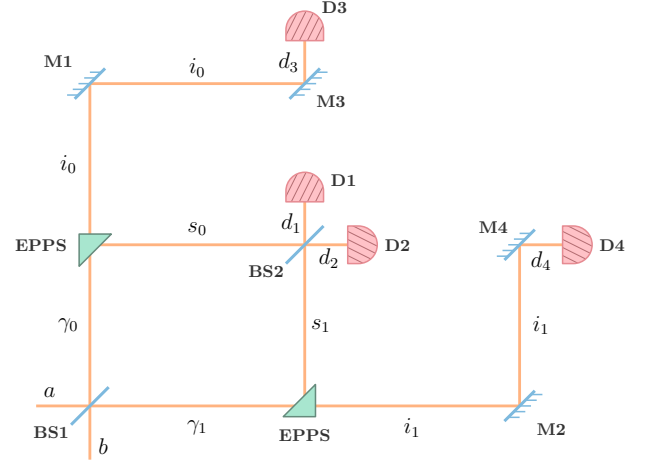

As often questioned in literature, ``What if we try to determine which arm the signal photon took in the signal MZI by detecting which path its entangled idler partner took in the idler MZI?'' Figure~\ref{fig:whichway} shows this setup. Here, mirrors $M_3$ and $M_4$ direct the idler modes $i_0$ and $i_1$ straight to detectors $D_3$ (mode $\ket{d_3}$) and $D_4$ (mode $\ket{d_4}$) in the idler MZI. 
The conventional argument given is that if the idler photon is detected at $D_3$ ($D_4$), from Equation~\ref{eq:entangled_initial} we know that the corresponding idler photon must have come through the $i_0$ ($i_1$) arm. This means that detections at $D_3$ and $D_4$ reveal which path the signal photon took. The final state of the setup is given by (see Appendix I):
\begin{equation}
\ket{\psi_{\text{II}}} = -\frac{1}{2} \Big[ \ket{\overline{D_3}}\left(i\ket{\overline{D_1}} + \ket{\overline{D_2}}\right) + \ket{\overline{D_4}}\left(\ket{\overline{D_1}} + i\ket{\overline{D_2}}\right) \Big].
\label{eq:psi_final_fig4}
\end{equation}
Because we now have which-path information, the coincidence count probability conditioned on either $D_3$ or $D_4$ shows that the signal MZI has equal probabilities of detection at $D_1$ and $D_2$. As a result, no interference is seen between $D_1$ and $D_2$; both detectors click with equal probability across the ensemble, as shown in Equation~\eqref{eq:psi_final_fig4}. In this case, the photon behaves like a particle.

\subsection{Case III}
\begin{figure}[htbp]
    \centering        
        \resizebox{\linewidth}{!}{\begin{tikzpicture}

\node[text=black, font=\large\bfseries] at (-0.75, 0.25) {$a$};
\node[text=black, font=\large\bfseries] at (0.25, -0.75) {$b$};

\node[text=black, font=\large\bfseries] at (2, -0.4) {$\gamma_1$};
\node[text=black, font=\large\bfseries] at (-0.4, 1.5) {$\gamma_0$};

\node[text=black, font=\large\bfseries] at (-0.4, 4.5) {$i_0$};
\node[text=black, font=\large\bfseries] at (2, 6.3) {$i_0$};
\node[text=black, font=\large\bfseries] at (5.3, 6.3) {$i_{0}^{'}$};

\node[text=black, font=\large\bfseries] at (8.4, 1.5) {$i_1$};
\node[text=black, font=\large\bfseries] at (6, -0.4) {$i_1$};
\node[text=black, font=\large\bfseries] at (8.4, 4.5) {$i_{1}^{'}$};

\node[text=black, font=\large\bfseries] at (2, 3.3) {$s_0$};
\node[text=black, font=\large\bfseries] at (4.3, 1.5) {$s_1$};

\node[text=black, font=\large\bfseries] at (3.75, 3.4) {$d_1$};
\node[text=black, font=\large\bfseries] at (4.45, 2.75) {$d_2$};

\node[text=black, font=\large\bfseries] at (3.7, 6.4) {$d_3$};
\node[text=black, font=\large\bfseries] at (8.4, 2.7) {$d_4$};

\node[text=black, font=\large\bfseries] at (7.7, 6.4) {$d_5$};
\node[text=black, font=\large\bfseries] at (8.4, 5.7) {$d_6$};

\drawbeam{-1}{0}{0}{0}{1.5pt}    
\drawbeam{ 0}{-1}{0}{0}{1.5pt}    
\drawbeam{0}{0}{8}{0}{1.5pt}       
\drawbeam{0}{0}{0}{6}{1.5pt}       
\drawbeam{0}{3}{5}{3}{1.5pt}       
\drawbeam{4}{0}{4}{4}{1.5pt}     
\drawbeam{0}{6}{9}{6}{1.5pt}      
\drawbeam{8}{0}{8}{7}{1.5pt}     
\drawbeam{8}{3}{9}{3}{1.5pt}      
\drawbeam{4}{6}{4}{7}{1.5pt}     

\drawbeamsplitter{0}{0}{0}{1}{125,185,222}
\node[below left, text=black!80, inner sep=8pt] at (-0.1,-0.1) {BS1};

\drawbeamsplitter{4}{3}{0}{1}{125,185,222}
\node[below left, text=black!80, inner sep=8pt] at (3.9,2.9) {BS2};

\drawbeamsplitter{4}{6}{0}{1}{125,185,222}
\node[below left, text=black!80, inner sep=8pt] at (3.9,5.9) {BS3};

\drawbeamsplitter{8}{3}{0}{1}{125,185,222}
\node[below left, text=black!80, inner sep=8pt] at (7.9,2.9) {BS4};

\drawbeamsplitter{8}{6}{0}{1}{125,185,222}
\node[below left, text=black!80, inner sep=8pt] at (7.9,5.9) {BS5};

\drawmirror{0}{6}{45}{1}{125,185,222}
\node[above left, text=black!80, inner sep=6pt] at (-0.1,6.1) {M1};

\drawmirror{8}{0}{225}{1}{125,185,222}
\node[below right, text=black!80, inner sep=6pt] at (8.1,-0.1) {M2};

\drawbbo{0.1}{2.9}{270}{1.2}{168,230,207}
\node[left, text=black!80, inner sep=12pt] at (0,3) {EPPS};

\drawbbo{3.9}{0.1}{90}{1.2}{168,230,207}
\node[below, text=black!80, inner sep=12pt] at (4,0) {EPPS};

\drawphase{6.5}{6}{0}{1}{220,220,220}
\node[above, text=black!80, inner sep=12pt] at (6.5,6) {$\phi$};

\drawdetector{4}{4}{0}{1}{255,196,163}
\node[right, text=black!80, inner sep=8pt] at (4.2,4.2) {D1};

\drawdetector{5}{3}{-90}{1}{255,196,163}
\node[right, text=black!80, inner sep=8pt] at (5.2,3) {D2};

\drawdetector{4}{7}{0}{1}{255,196,163}
\node[above right, text=black!80, inner sep=6pt] at (4.1,7.1) {D3};

\drawdetector{9}{3}{-90}{1}{255,196,163}
\node[right, text=black!80, inner sep=8pt] at (9.2,3) {D4};

\drawdetector{8}{7}{0}{1}{255,196,163}
\node[above right, text=black!80, inner sep=6pt] at (8.1,7.1) {D5};

\drawdetector{9}{6}{-90}{1}{255,196,163}
\node[right, text=black!80, inner sep=8pt] at (9.2,6) {D6};

\end{tikzpicture}}
    \caption{Extended MZI obtained by replacing mirrors $M_3$ and $M_4$ in the case II with 50:50 beam splitters $\text{BS}_3$ and $\text{BS}_4$ unifies both operational modes (case I and case II) into a single schematic. Reflected idler modes ($\ket{d_3}, \ket{d_4}$) yield which-path measurements at detectors $D_3$ and $D_4$, while transmitted modes ($\ket{i_0'}, \ket{i_1'}$) recombine at $\text{BS}_5$ for erasure measurements at $D_5$ and $D_6$. Arbitrarily extending the idler optical path lengths ensures signal registration at $D_1/D_2$ strictly precedes the idler measurement choice.}
    \label{fig:delayedchoice}
\end{figure}
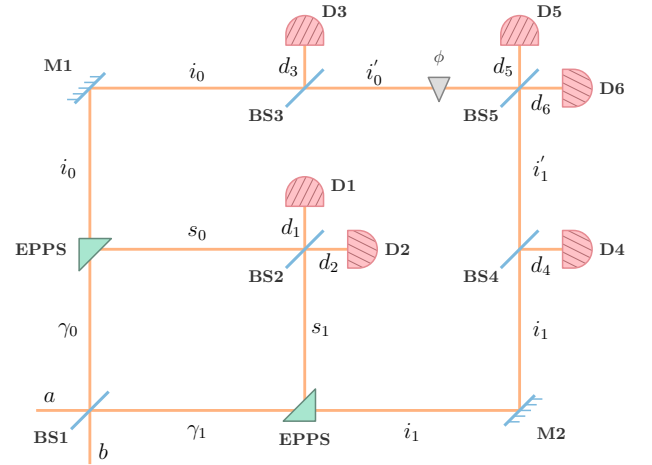

In the third configuration, we combine the first two set-ups. We replace mirrors $M_3$ and $M_4$ with $50:50$ beam splitters, called $\text{BS}_3$ and $\text{BS}_4$, as shown in Figure \ref{fig:delayedchoice}. To implement a delayed choice, we make the idler arms $i_0$ and $i_1$ much {longer} than the signal arms $s_0$ and $s_1$. This way, the signal photons are detected at $D_1$ or $D_2$ before the idler photons reach their beam splitters. This setup  corresponds to the delayed-choice quantum eraser (DCQE). The final state, after all transformations, is given by (see Appendix I):
\begin{eqnarray}
\ket{\psi_{\text{III}}} =& -\frac{1}{2\sqrt{2}} \Big[ \ket{\overline{D_3}} \Big( i\ket{\overline{D_1}} + \ket{\overline{D_2}} \Big)+\ket{\overline{D_4}} \Big( \ket{\overline{D_1}} + i\ket{\overline{D_2}} \Big)\Big]\nonumber\\
&+i\frac{1}{2}\Big[\ket{\overline{D_5}}\ket{\overline{D_1}} -\ket{\overline{D_6}}\ket{\overline{D_2}}\Big].~\label{eq:psi_final_fig5}
\end{eqnarray}
Equation~\eqref{eq:psi_final_fig5} encodes the complete joint measurement statistics across the six-detectors.
The detector operations fall into two complementary sets. When an idler photon is reflected at $\text{BS}_3$ (triggering $D_3$) or $\text{BS}_4$ (triggering $D_4$), the path information of the signal photon is said to be known. A $D_3$ click isolates signal photons originating from path $s_0$, while a $D_4$ click isolates path $s_1$. In both cases, the conditional signal distribution at $D_1$ and $D_2$ shows no  interference. Conversely, beam splitter $\text{BS}_5$ is said to erase the path record by coherently mixing modes $i_0'$ and $i_1'$. As in the setup by Kim et al. \cite{kim2000}, a coincidence click between $D_2$ and $D_6$ isolates the sub-ensemble corresponding to the balanced MZI \textit{interference}. Similarly, a coincidence click between $D_1$ and $D_5$ corresponds to the \textit{complementary-interference}.

The choice of whether an idler photon is registered at $D_3/ D_4$ or $D_5/ D_6$ is determined long after the signal photon has registered at $D_1$ or $D_2$. The sub-ensembles corresponding to \textit{interference} ($D_6$ clicking), \textit{complementary-interference} ($D_5$ clicking), and which-path knowledge ($D_3$ or $D_4$ clicking) when combined together yield \emph{no overall interference}, demonstrating that erasure is realized strictly via joint post-selection.

The physical outcomes encoded in Equation~\eqref{eq:psi_final_fig5} lend themselves to two distinct conceptual interpretations, highlighting the distinction between classical teleological reasoning and formal quantum mechanics. If one enforces a strict chronological narrative where the signal photon's registration at $D_1$ or $D_2$ precedes the idler photon's arrival at $D_3$-$D_6$, an apparent paradox emerges. One is tempted to ask how the earlier signal detection knows whether its idler partner will subsequently be measured in which-path detectors ($D_3/D_4$) or an erasure detectors ($D_5/D_6$). Formulating the question in this manner forces the conclusion that the future idler detection retroactively dictates the past behavior of the signal photon. Standard quantum mechanics resolves this pseudo-paradox without invoking past-directed influences. The routing of the idler photon toward $D_3/ D_4$ versus $D_5/ D_6$ simply corresponds to performing a measurement in one of two mutually unbiased measurement bases \cite{chiou2023delayedchoice}. The non-separable joint state established in Eq.~\ref{eq:entangled_initial} dictates the correlations, and the future idler detections act merely as a filtering key for data post-selection. To make the underlying physics easier to understand within a standard quantum mechanics framework, we map the delayed-choice protocol onto a familiar Stern-Gerlach apparatus measuring spin-$1/2$ entangled pairs, which we discuss in the next section.


\section*{A Stern–Gerlach Analogy for Delayed Choice Quantum Eraser  }

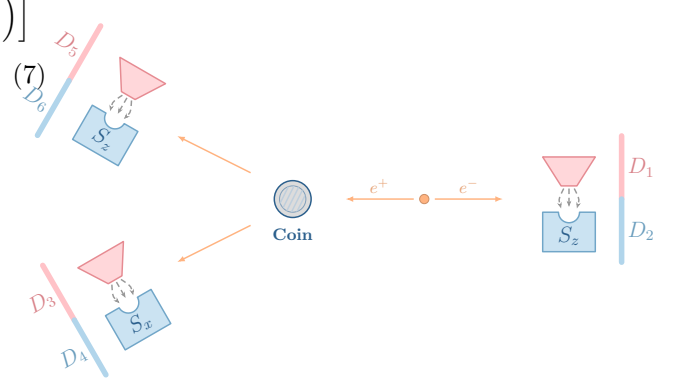
\begin{figure}[htbp]
    \centering

        \resizebox{\linewidth}{!}{\begin{tikzpicture}

\drawSG{1.25}{1.75}{-30}{1}{$S_z$}
\drawSG{1.25}{-1.75}{30}{1}{$S_x$}
\drawSG{9.75}{0}{0}{1}{$S_z$}

\DetectorPlate{0.25}{2.25}{-30}{1}{$D_5$}{$D_6$}{left}
\DetectorPlate{0.35}{-2.3}{30}{1}{$D_3$}{$D_4$}{left}
\DetectorPlate{10.75}{0}{0}{1}{$D_1$}{$D_2$}{right}


\begin{scope}[shift={(4.5, 0)}]
    \filldraw[fill=pastelgray, draw=textblue, thick] (0,0) circle (0.35);
    \draw[textblue!60, thin] (0,0) circle (0.25);
    
    \begin{scope}
        \clip (0,0) circle (0.25);
        \foreach \x in {-0.4,-0.3,...,0.4} {
            \draw[textblue!30, thick] (\x, -0.4) -- (\x+0.4, 0.4);
        }
    \end{scope}
    
    \node[below=0.45cm] at (0, 0) {\textsc{Coin}};
\end{scope}

\begin{scope}[shift={(7, 0)}]
    \filldraw[fill=pastelorange, draw=pastelorange!80!black] (0,0) circle (2.5pt);
    
    \draw[pastelorange, thick, -{Latex[length=1.5mm]}] (-0.2, 0) -- (-1.5, 0) node[above, midway, text=pastelorange!90!black] {$e^+$};

        \draw[pastelorange, thick, -{Latex[length=1.5mm]}] (-3.3, 0.5) -- (-4.7, 1.2) node[above, midway, text=pastelorange!90!black] {};

        \draw[pastelorange, thick, -{Latex[length=1.5mm]}] (-3.3, -0.5) -- (-4.7, -1.2) node[above, midway, text=pastelorange!90!black] {};
    
    \draw[pastelorange, thick, -{Latex[length=1.5mm]}] (0.2, 0) -- (1.5, 0) node[above, midway, text=pastelorange!90!black] {$e^-$};
\end{scope}

\end{tikzpicture}}
    \caption{Schematic diagram of a two-way Stern-Gerlach setup. A central source emits a spin-entangled electron-positron pair ($e^-$, $e^+$). The electron's spin is measured by a Stern-Gerlach apparatus along the z-axis ($S_z$), while the positron's measurement basis is dynamically selected between the x-axis ($S_x$) and z-axis ($S_z$) using a quantum coin.
}
    \label{fig:sterngerlach}
\end{figure}

As shown in Figure~\ref{fig:sterngerlach}, an electron-positron pair ($e^+$ moving left and $e^-$ moving right) is prepared in the following maximally entangled spin state (by applying localized unitary operations to one of the particles in the singlet state after the decay~\cite{nielsen2000quantum}):
\begin{equation}
\ket{\psi} = \frac{1}{\sqrt{2}}\left( \ket{0}\ket{0} + \ket{1}\ket{1} \right),
\label{eq:sg_initial_state}
\end{equation}
where $\ket{0} \equiv \ket{\uparrow_z}$ and $\ket{1} \equiv \ket{\downarrow_z}$ are the eigenstates of the Pauli spin operator $\hat{\sigma}_z$.  In this notation, the first ket refers to the state of positron and the second ket refers to the state of electron. The electron travels toward a $SG_z$ apparatus on the right side of the setup and is measured in the $z$-basis ($\ket{0}, \ket{1}$). The positron travels a much longer distance to the left toward a quantum device, which we call a quantum coin, prepared in the state $\frac{1}{\sqrt{2}}(\ket{H}+\ket{T})$. If the quantum coin gives Tails ($\ket{T}$), the positron is sent to a $SG_z$ apparatus and measured in the basis $\{\ket{0}, \ket{1}\}$. If it gives Heads ($\ket{H}$), the positron is sent to a $SG_x$ apparatus and measured in the basis $\{\ket{+}, \ket{-}\}$, where $\ket{\pm} = \frac{1}{\sqrt{2}}(\ket{0} \pm \ket{1})$. Because the positron travels a longer path than the electron, the electron is measured before the positron.

The detection events can be summarized as follows: For electrons, a collapse into $\ket{0}$ at detector plate $D_1$ corresponds to detector $\ket{\overline{D_1}}$ clicking in the MZI, while a collapse into $\ket{1}$ at plate $D_2$ corresponds to detector $\ket{\overline{D_2}}$ clicking. For positrons, the mapping depends on the measurement basis. If the positron collapses into $\ket{0}$ at plate $D_5$ or into $\ket{1}$ at plate $D_6$, the corresponding MZI detectors are $\ket{\overline{D_5}}$ and $\ket{\overline{D_6}}$, respectively. However, if the positron collapses into $\ket{+}$ at plate $D_3$ or into $\ket{-}$ at plate $D_4$, then the corresponding MZI detectors are instead $\ket{\overline{D_3}}$ and $\ket{\overline{D_4}}$, respectively.

Suppose the electron collapses in $\ket{0}$ , then due to the non-separable correlations established earlier, the total joint state collapses into:
\begin{equation}
\ket{\psi_{\text{collapsed}}} = \ket{0}\ket{0}.
\end{equation}
Even though the positron has not yet reached the detector plates, its state is conditioned in the $z$-basis. The subsequent outcome of which detector plate out of $D_3, D_4, D_5$ or $D_6$ the positron hits depends entirely on the choice of measurement basis for positron. If the coin lands tails and positron moves towards $SG_z$,  the positron strikes at plate $D_5$ ($\ket{0}$) with certainty, while $D_6$ ($\ket{1}$) is stuck with no probability. This recovers the one-to-one coincidence matching $(D_1, D_5)$, precisely as predicted for the case III extended MZI configuration. If the coin lands heads and positron travels towards $SG_x$, the positron's state must be re-expressed in the $x$-basis:
    \begin{equation}
    \ket{\psi_{\rm collapsed}} = \frac{1}{\sqrt{2}}\left( \ket{+} + \ket{-}\right)\ket{0}.
    \end{equation}
Projection onto $SG_x$ yields equal probabilities of striking detector plate $D_3$ ($\ket{+}$) or  $D_4$ ($\ket{-}$). The algebraic equivalence between this SG model and the case III extended MZI setup demonstrates that the delayed-choice phenomenology is nothing more than standard quantum mechanical basis change. Note that this replica does not strictly implement a which-path or quantum eraser experiment. Instead, it simply reproduces the algebraic results obtained in case III of the extended MZI. This simplified setup helps us understand the controversial conditional probability data from the delayed-choice experiment using standard quantum mechanics and basis changes.

Asking how the electron striking at $D_1$ knew whether the positron would land at $D_5$ versus $D_3/D_4$, even before positron reached detector plates, is an illegitimate question. The electron's initial detection fixes a state that yields matching standard quantum mechanical predictions once a specific measurement basis ($\hat{\sigma}_z$ or $\hat{\sigma}_x$) is chosen for the positron. No retrocausal mechanism or advanced signaling is required; the observed statistics follow directly from elementary quantum state reduction over an entangled pair.  Similarly, in case III of the extended MZI, the apparent paradox is easily resolved within standard quantum mechanics. The conceptual confusion arises from the fact that the photon's state is entangled with its spatial path, prompting the question of which path the photon took and leading us to misinterpret the action of beam splitter $\text{BS}_5$ as erasing which-path information. In reality, before collapse occurs, assigning a definite path to the particle constitutes an unjustified ontological commitment; the particle simply cannot be said to have taken one path or the other. Whether a collapse occurs at $D_3/D_4$ or $D_5/D_6$ simply reflects the choice of measurement basis. Hence, nothing is physically erased or delayed; the protocol merely corresponds to a basis rotation. As pointed out by Kastner \cite{kastner2019delayed}, our ability to express a state in different bases neither adds nor removes information, nor does it warrant labeling a quantum state as inherently containing or erasing path information.

\section*{Game-theoretic analogy}
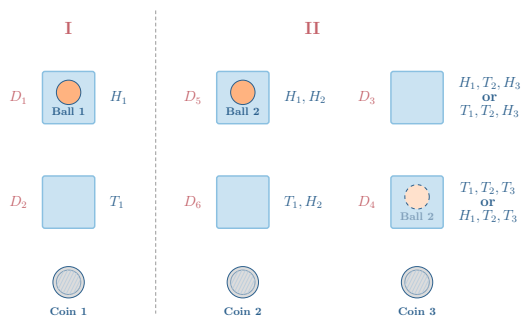
\begin{figure}[htbp]
    \centering
        \resizebox{0.8\linewidth}{!}{\begin{tikzpicture}



\node[redlabel] at (0, 5) {I};

\node[box] (D1) at (0, 3) {};
\node[left=0.3cm of D1, dlabel] {$D_1$};
\node[right=0.3cm of D1, greenlabel] {$H_1$};
\filldraw[fill=pastelorange!100, draw=textblue!100, thick] ($(D1.center)+(0,0.15)$) circle (0.35cm);
\node[balltext] at ($(D1.center)+(0,-0.4)$) {Ball 1};

\node[box] (D2) at (0, 0) {};
\node[left=0.3cm of D2, dlabel] {$D_2$};
\node[right=0.3cm of D2, greenlabel] {$T_1$};

\drawCoin{0}{-2.3}{\textsc{Coin 1}}

\draw[gray!80, densely dashed, thick] (2.5, 5.5) -- (2.5, -3.2);


\node[redlabel] at (7, 5) {II};


\node[box] (D3) at (5, 3) {};
\node[left=0.3cm of D3, dlabel] {$D_5$};
\node[right=0.3cm of D3, greenlabel] {$H_1, H_2$};
\filldraw[fill=pastelorange!100, draw=textblue!100, thick] ($(D3.center)+(0,0.15)$) circle (0.35cm);
\node[balltext] at ($(D3.center)+(0,-0.4)$) {Ball 2};

\node[box] (D4) at (5, 0) {};
\node[left=0.3cm of D4, dlabel] {$D_6$};
\node[right=0.3cm of D4, greenlabel] {$T_1, H_2$};

\drawCoin{5}{-2.3}{\textsc{Coin 2}}


\node[box] (D5) at (10, 3) {};
\node[left=0.3cm of D5, dlabel] {$D_3$};
\node[right=0.3cm of D5, greenlabel, align=center] {$H_1, T_2, H_3$ \\[-0.5ex] or \\[-0.5ex] $T_1, T_2, H_3$};

\node[box] (D6) at (10, 0) {};
\node[left=0.3cm of D6, dlabel] {$D_4$};
\node[right=0.3cm of D6, greenlabel, align=center] {$T_1, T_2, T_3$ \\[-0.5ex] or \\[-0.5ex] $H_1, T_2, T_3$};

\filldraw[fill=pastelorange!40, draw=textblue, thick, dashed] ($(D6.center)+(0,0.15)$) circle (0.35cm);
\node[balltext, text=textblue!60] at ($(D6.center)+(0,-0.4)$) {Ball 2};

\drawCoin{10}{-2.3}{\textsc{Coin 3}}


\end{tikzpicture}}
    \caption{ Schematic of the ball-and-box game used to model the delayed-choice quantum eraser. Alice initializes the system via random coin flips, and Bob examines the final ball distribution. The resulting correlations mimic those of the quantum eraser, demonstrating that post-selection, not retrocausality, generates the observed patterns.}
    \label{fig:coinmodel}
\end{figure}

To demonstrate that the phenomenology of the delayed-choice quantum eraser does not necessitate retrocausal information transfer, we can model the setup through a game-theoretic lens using balls and a stochastic sorting mechanism like coin flips.
Consider a game played by Alice and Bob. Alice's objective is to prepare a specific arrangement of balls in boxes, while Bob's role is to analyze her final arrangement and map out the statistical correlations.
\\
\textit{\emph{Alice's setup}:}
Alice uses two balls (Ball 1 and Ball 2) and six boxes ($D_1$ through $D_6$). She determines their placement through a sequence of independent coin flips:
Alice flips the first coin. As shown in Fig.~\ref{fig:coinmodel}, if Heads ($H_1$): Ball 1 is placed in $D_1$ and Ball 2 in $D_5$. If Tails ($T_1$): Ball 1 is placed in $D_2$ and Ball 2 in $D_6$. 
Alice flips a second coin to determine what happens to Ball 2. If Heads ($H_2$): Ball 2 remains undisturbed in its assigned box ($D_5$ or $D_6$). If Tails ($T_2$): Ball 2 is subjected to a final sorting. 
If Coin 2 was Tails, Alice flips a third coin to redirect Ball 2. If Heads ($H_3$): Ball 2 is moved to $D_3$. If Tails ($T_3$): Ball 2 is moved to $D_4$. Alice compiles the positions of Ball 1 and Ball 2 in all the boxes and gives this data to Bob.
\\
 \textit{\emph{Bob's analysis \& the paradox}:}
Bob takes over to analyze the data for placement of Ball~2 relative to Ball~1. He knows that Ball 1 was placed before placing Ball 2. He discovers a striking pattern: if Ball~2 lands in~$D_5$, Ball~1 is always found in~$D_1$; if Ball~2 lands in~$D_6$, Ball~1 is always in~$D_2$. However, when Ball~2 ends up in~$D_3$ or~$D_4$, the outcome for Ball~1 becomes evenly split, it appears in~$D_1$ and~$D_2$ with equal probability.
This leads Bob to a natural question: How could Ball~1 ``know'' in advance whether it needed to land definitively in~$D_1$ or~$D_2$, or whether it should instead be distributed evenly between the two? %
\\\\
\textit{\emph{The resolution}:}
The answer is that Ball 1 requires no such foresight. Its position was permanently fixed by Coin 1. Coins 2 and 3 simply acted as a delayed sorting mechanism for Ball 2. The mistake lies in treating Bob's final, sorted sub-ensembles as if they actively caused the past.
This perfectly captures the operational logic of the delayed-choice quantum eraser, where Balls 1 and 2 represent the entangled signal and idler photons. Demanding to know how the signal photon ``knew" the fate of its idler counterpart is an interpretive artifact. It arises from trying to construct a backward-moving chain of cause-and-effect out of post-selected data. 

This game provides a practical framework for teaching the well-established epistemological distinction between causation and correlation. The game implies backwards correlation yet it still rules out backward causation and avoids any conflict with relativity(see \cite{gaasbeek2010demystifying}).

\section{Conclusion}
In this paper, we have addressed the long-standing debate surrounding the delayed-choice quantum eraser by showing that its apparent paradoxes can be fully resolved using standard quantum mechanics. In our extended Mach–Zehnder interferometer, designed to match every operational feature of the original Kim et al. experiment~\cite{kim2000}, we find that the choice between which-path and erasure detectors is simply a selection between different mutually unbiased measurement bases. Translating the extended MZI setup into a two-way Stern-Gerlach framework makes the underlying quantum mechanics significantly more transparent.
Finally, our pedagogical game illustrates that demanding a backward-in-time cause-and-effect narrative for post-selected sub-ensembles is an interpretive mistake rather than a physical mystery. More generally, this work underscores that, while quantum mechanics is counterintuitive, not every counterintuitive effect represents its true nature. Some paradoxes, including retrocausal erasure, are simply conceptual confusions arising as a byproduct of quantum mechanics being counterintuitive in the first place, creating a self-reinforcing loop of misunderstanding. 

\section*{Acknowledgment}
We thank Department of Science and Technology, India for the FIST program to the Physics Department, GNDU.

\bibliography{main} 

\begin{thebibliography}{31}%
\makeatletter
\providecommand \@ifxundefined [1]{%
 \@ifx{#1\undefined}
}%
\providecommand \@ifnum [1]{%
 \ifnum #1\expandafter \@firstoftwo
 \else \expandafter \@secondoftwo
 \fi
}%
\providecommand \@ifx [1]{%
 \ifx #1\expandafter \@firstoftwo
 \else \expandafter \@secondoftwo
 \fi
}%
\providecommand \natexlab [1]{#1}%
\providecommand \enquote  [1]{``#1''}%
\providecommand \bibnamefont  [1]{#1}%
\providecommand \bibfnamefont [1]{#1}%
\providecommand \citenamefont [1]{#1}%
\providecommand \href@noop [0]{\@secondoftwo}%
\providecommand \href [0]{\begingroup \@sanitize@url \@href}%
\providecommand \@href[1]{\@@startlink{#1}\@@href}%
\providecommand \@@href[1]{\endgroup#1\@@endlink}%
\providecommand \@sanitize@url [0]{\catcode `\\12\catcode `\$12\catcode `\&12\catcode `\#12\catcode `\^12\catcode `\_12\catcode `\%12\relax}%
\providecommand \@@startlink[1]{}%
\providecommand \@@endlink[0]{}%
\providecommand \url  [0]{\begingroup\@sanitize@url \@url }%
\providecommand \@url [1]{\endgroup\@href {#1}{\urlprefix }}%
\providecommand \urlprefix  [0]{URL }%
\providecommand \Eprint [0]{\href }%
\providecommand \doibase [0]{https://doi.org/}%
\providecommand \selectlanguage [0]{\@gobble}%
\providecommand \bibinfo  [0]{\@secondoftwo}%
\providecommand \bibfield  [0]{\@secondoftwo}%
\providecommand \translation [1]{[#1]}%
\providecommand \BibitemOpen [0]{}%
\providecommand \bibitemStop [0]{}%
\providecommand \bibitemNoStop [0]{.\EOS\space}%
\providecommand \EOS [0]{\spacefactor3000\relax}%
\providecommand \BibitemShut  [1]{\csname bibitem#1\endcsname}%
\let\auto@bib@innerbib\@empty
\bibitem [{\citenamefont {Kim}\ \emph {et~al.}(2000)\citenamefont {Kim}, \citenamefont {Yu}, \citenamefont {Kulik}, \citenamefont {Shih},\ and\ \citenamefont {Scully}}]{kim2000}%
  \BibitemOpen
  \bibfield  {author} {\bibinfo {author} {\bibfnamefont {Y.-H.}\ \bibnamefont {Kim}}, \bibinfo {author} {\bibfnamefont {R.}~\bibnamefont {Yu}}, \bibinfo {author} {\bibfnamefont {S.~P.}\ \bibnamefont {Kulik}}, \bibinfo {author} {\bibfnamefont {Y.}~\bibnamefont {Shih}},\ and\ \bibinfo {author} {\bibfnamefont {M.~O.}\ \bibnamefont {Scully}},\ }\bibfield  {title} {\bibinfo {title} {Delayed ``choice'' quantum eraser},\ }\href {https://doi.org/10.1103/PhysRevLett.84.1} {\bibfield  {journal} {\bibinfo  {journal} {Physical Review Letters}\ }\textbf {\bibinfo {volume} {84}},\ \bibinfo {pages} {1} (\bibinfo {year} {2000})}\BibitemShut {NoStop}%
\bibitem [{\citenamefont {Wheeler}(1978)}]{wheeler1978}%
  \BibitemOpen
  \bibfield  {author} {\bibinfo {author} {\bibfnamefont {J.~A.}\ \bibnamefont {Wheeler}},\ }\bibfield  {title} {\bibinfo {title} {The ``past" and the ``delayed-choice" double-slit experiment},\ }in\ \href@noop {} {\emph {\bibinfo {booktitle} {Mathematical Foundations of Quantum Theory}}},\ \bibinfo {editor} {edited by\ \bibinfo {editor} {\bibfnamefont {A.~R.}\ \bibnamefont {Marlow}}}\ (\bibinfo  {publisher} {Academic Press},\ \bibinfo {address} {New York},\ \bibinfo {year} {1978})\ pp.\ \bibinfo {pages} {9--48}\BibitemShut {NoStop}%
\bibitem [{\citenamefont {Qureshi}(2020)}]{qureshi2020}%
  \BibitemOpen
  \bibfield  {author} {\bibinfo {author} {\bibfnamefont {T.}~\bibnamefont {Qureshi}},\ }\bibfield  {title} {\bibinfo {title} {Demystifying the delayed-choice quantum eraser},\ }\href {https://doi.org/10.1088/1361-6404/ab923e} {\bibfield  {journal} {\bibinfo  {journal} {European Journal of Physics}\ }\textbf {\bibinfo {volume} {41}},\ \bibinfo {pages} {055403} (\bibinfo {year} {2020})}\BibitemShut {NoStop}%
\bibitem [{\citenamefont {Qureshi}(2021)}]{qureshi2021delayedchoice}%
  \BibitemOpen
  \bibfield  {author} {\bibinfo {author} {\bibfnamefont {T.}~\bibnamefont {Qureshi}},\ }\bibfield  {title} {\bibinfo {title} {The delayed-choice quantum eraser leaves no choice},\ }\href {https://doi.org/10.1007/s10773-021-04906-w} {\bibfield  {journal} {\bibinfo  {journal} {International Journal of Theoretical Physics}\ }\textbf {\bibinfo {volume} {60}},\ \bibinfo {pages} {3076} (\bibinfo {year} {2021})}\BibitemShut {NoStop}%
\bibitem [{\citenamefont {Chiou}(2023)}]{chiou2023delayedchoice}%
  \BibitemOpen
  \bibfield  {author} {\bibinfo {author} {\bibfnamefont {D.-W.}\ \bibnamefont {Chiou}},\ }\bibfield  {title} {\bibinfo {title} {Delayed-choice quantum erasers and the einstein-podolsky-rosen paradox},\ }\bibfield  {journal} {\bibinfo  {journal} {International Journal of Theoretical Physics}\ }\textbf {\bibinfo {volume} {62}},\ \href {https://doi.org/10.1007/s10773-023-05370-4} {10.1007/s10773-023-05370-4} (\bibinfo {year} {2023})\BibitemShut {NoStop}%
\bibitem [{\citenamefont {Chaves}\ \emph {et~al.}(2018)\citenamefont {Chaves}, \citenamefont {Lemos},\ and\ \citenamefont {Pienaar}}]{chaves2018}%
  \BibitemOpen
  \bibfield  {author} {\bibinfo {author} {\bibfnamefont {R.}~\bibnamefont {Chaves}}, \bibinfo {author} {\bibfnamefont {G.~B.}\ \bibnamefont {Lemos}},\ and\ \bibinfo {author} {\bibfnamefont {J.}~\bibnamefont {Pienaar}},\ }\bibfield  {title} {\bibinfo {title} {Causal modeling the delayed-choice experiment},\ }\href {https://doi.org/10.1103/PhysRevLett.120.190401} {\bibfield  {journal} {\bibinfo  {journal} {Phys. Rev. Lett.}\ }\textbf {\bibinfo {volume} {120}},\ \bibinfo {pages} {190401} (\bibinfo {year} {2018})}\BibitemShut {NoStop}%
\bibitem [{\citenamefont {Bracken}\ \emph {et~al.}(2021)\citenamefont {Bracken}, \citenamefont {Hance},\ and\ \citenamefont {Hossenfelder}}]{sabine2021}%
  \BibitemOpen
  \bibfield  {author} {\bibinfo {author} {\bibfnamefont {C.}~\bibnamefont {Bracken}}, \bibinfo {author} {\bibfnamefont {J.~R.}\ \bibnamefont {Hance}},\ and\ \bibinfo {author} {\bibfnamefont {S.}~\bibnamefont {Hossenfelder}},\ }\href {https://doi.org/2111.09347} {\bibinfo {title} {The quantum eraser paradox}} (\bibinfo {year} {2021})\BibitemShut {NoStop}%
\bibitem [{\citenamefont {Walborn}\ \emph {et~al.}(2002{\natexlab{a}})\citenamefont {Walborn}, \citenamefont {Terra~Cunha}, \citenamefont {P\'adua},\ and\ \citenamefont {Monken}}]{exp1}%
  \BibitemOpen
  \bibfield  {author} {\bibinfo {author} {\bibfnamefont {S.~P.}\ \bibnamefont {Walborn}}, \bibinfo {author} {\bibfnamefont {M.~O.}\ \bibnamefont {Terra~Cunha}}, \bibinfo {author} {\bibfnamefont {S.}~\bibnamefont {P\'adua}},\ and\ \bibinfo {author} {\bibfnamefont {C.~H.}\ \bibnamefont {Monken}},\ }\bibfield  {title} {\bibinfo {title} {Double-slit quantum eraser},\ }\href {https://doi.org/10.1103/PhysRevA.65.033818} {\bibfield  {journal} {\bibinfo  {journal} {Phys. Rev. A}\ }\textbf {\bibinfo {volume} {65}},\ \bibinfo {pages} {033818} (\bibinfo {year} {2002}{\natexlab{a}})}\BibitemShut {NoStop}%
\bibitem [{\citenamefont {Ma}\ \emph {et~al.}(2013)\citenamefont {Ma}, \citenamefont {Kofler}, \citenamefont {Qarry}, \citenamefont {Tetik}, \citenamefont {Scheidl}, \citenamefont {Ursin}, \citenamefont {Ramelow}, \citenamefont {Herbst}, \citenamefont {Ratschbacher}, \citenamefont {Fedrizzi}, \citenamefont {Jennewein},\ and\ \citenamefont {Zeilinger}}]{exp2}%
  \BibitemOpen
  \bibfield  {author} {\bibinfo {author} {\bibfnamefont {X.-S.}\ \bibnamefont {Ma}}, \bibinfo {author} {\bibfnamefont {J.}~\bibnamefont {Kofler}}, \bibinfo {author} {\bibfnamefont {A.}~\bibnamefont {Qarry}}, \bibinfo {author} {\bibfnamefont {N.}~\bibnamefont {Tetik}}, \bibinfo {author} {\bibfnamefont {T.}~\bibnamefont {Scheidl}}, \bibinfo {author} {\bibfnamefont {R.}~\bibnamefont {Ursin}}, \bibinfo {author} {\bibfnamefont {S.}~\bibnamefont {Ramelow}}, \bibinfo {author} {\bibfnamefont {T.}~\bibnamefont {Herbst}}, \bibinfo {author} {\bibfnamefont {L.}~\bibnamefont {Ratschbacher}}, \bibinfo {author} {\bibfnamefont {A.}~\bibnamefont {Fedrizzi}}, \bibinfo {author} {\bibfnamefont {T.}~\bibnamefont {Jennewein}},\ and\ \bibinfo {author} {\bibfnamefont {A.}~\bibnamefont {Zeilinger}},\ }\bibfield  {title} {\bibinfo {title} {Quantum erasure with causally disconnected choice},\ }\href {https://doi.org/10.1073/pnas.1213201110} {\bibfield  {journal} {\bibinfo  {journal} {Proceedings of the National Academy of Sciences}\
  }\textbf {\bibinfo {volume} {110}},\ \bibinfo {pages} {1221} (\bibinfo {year} {2013})}\BibitemShut {NoStop}%
\bibitem [{\citenamefont {Kastner}(2019)}]{kastner2019delayed}%
  \BibitemOpen
  \bibfield  {author} {\bibinfo {author} {\bibfnamefont {R.~E.}\ \bibnamefont {Kastner}},\ }\bibfield  {title} {\bibinfo {title} {The ‘delayed choice quantum eraser’ neither erases nor delays},\ }\href {https://doi.org/10.1007/s10701-019-00278-8} {\bibfield  {journal} {\bibinfo  {journal} {Foundations of Physics}\ }\textbf {\bibinfo {volume} {49}},\ \bibinfo {pages} {717} (\bibinfo {year} {2019})}\BibitemShut {NoStop}%
\bibitem [{\citenamefont {Aharonov}\ and\ \citenamefont {Zubairy}(2005)}]{aharonov2005}%
  \BibitemOpen
  \bibfield  {author} {\bibinfo {author} {\bibfnamefont {Y.}~\bibnamefont {Aharonov}}\ and\ \bibinfo {author} {\bibfnamefont {M.~S.}\ \bibnamefont {Zubairy}},\ }\bibfield  {title} {\bibinfo {title} {Time and the quantum: Erasing the past and impacting the future},\ }\href {https://doi.org/10.1126/science.1107787} {\bibfield  {journal} {\bibinfo  {journal} {Science}\ }\textbf {\bibinfo {volume} {307}},\ \bibinfo {pages} {875} (\bibinfo {year} {2005})},\ \Eprint {https://arxiv.org/abs/https://www.science.org/doi/pdf/10.1126/science.1107787} {https://www.science.org/doi/pdf/10.1126/science.1107787} \BibitemShut {NoStop}%
\bibitem [{\citenamefont {Hiley}\ and\ \citenamefont {Callaghan}(2006)}]{hiley2006}%
  \BibitemOpen
  \bibfield  {author} {\bibinfo {author} {\bibfnamefont {B.~J.}\ \bibnamefont {Hiley}}\ and\ \bibinfo {author} {\bibfnamefont {R.~E.}\ \bibnamefont {Callaghan}},\ }\bibfield  {title} {\bibinfo {title} {What is erased in the quantum erasure?},\ }\href {https://doi.org/10.1007/s10701-006-9086-4} {\bibfield  {journal} {\bibinfo  {journal} {Foundations of Physics}\ }\textbf {\bibinfo {volume} {36}},\ \bibinfo {pages} {1869} (\bibinfo {year} {2006})}\BibitemShut {NoStop}%
\bibitem [{\citenamefont {Ellerman}(2015)}]{ellerman2015}%
  \BibitemOpen
  \bibfield  {author} {\bibinfo {author} {\bibfnamefont {D.~P.}\ \bibnamefont {Ellerman}},\ }\bibfield  {title} {\bibinfo {title} {Why delayed choice experiments do not imply retrocausality},\ }\href {https://doi.org/https://doi.org/10.1007/s40509-014-0026-2} {\bibfield  {journal} {\bibinfo  {journal} {Quantum Studies: Mathematics and Foundations}\ }\textbf {\bibinfo {volume} {2}},\ \bibinfo {pages} {183 } (\bibinfo {year} {2015})}\BibitemShut {NoStop}%
\bibitem [{\citenamefont {Fankhauser}(2017)}]{fankhauser2017}%
  \BibitemOpen
  \bibfield  {author} {\bibinfo {author} {\bibfnamefont {J.}~\bibnamefont {Fankhauser}},\ }\bibfield  {title} {\bibinfo {title} {Taming the delayed choice quantum eraser},\ }\href {https://api.semanticscholar.org/CorpusID:53574007} {\bibfield  {journal} {\bibinfo  {journal} {Quanta}\ } (\bibinfo {year} {2017})}\BibitemShut {NoStop}%
\bibitem [{\citenamefont {{Englert}}\ \emph {et~al.}(1999)\citenamefont {{Englert}}, \citenamefont {{Scully}},\ and\ \citenamefont {{Walther}}}]{englert1999}%
  \BibitemOpen
  \bibfield  {author} {\bibinfo {author} {\bibfnamefont {B.-G.}\ \bibnamefont {{Englert}}}, \bibinfo {author} {\bibfnamefont {M.~O.}\ \bibnamefont {{Scully}}},\ and\ \bibinfo {author} {\bibfnamefont {H.}~\bibnamefont {{Walther}}},\ }\bibfield  {title} {\bibinfo {title} {{Quantum erasure in double-slit interferometers with which-way detectors}},\ }\href {https://doi.org/10.1119/1.19257} {\bibfield  {journal} {\bibinfo  {journal} {American Journal of Physics}\ }\textbf {\bibinfo {volume} {67}},\ \bibinfo {pages} {325} (\bibinfo {year} {1999})}\BibitemShut {NoStop}%
\bibitem [{\citenamefont {Mohrhoff}(1999)}]{mohrhoff1999}%
  \BibitemOpen
  \bibfield  {author} {\bibinfo {author} {\bibfnamefont {U.~J.}\ \bibnamefont {Mohrhoff}},\ }\bibfield  {title} {\bibinfo {title} {Objectivity, retrocausation, and the experiment of englert, scully, and walther},\ }\href {https://api.semanticscholar.org/CorpusID:121734504} {\bibfield  {journal} {\bibinfo  {journal} {American Journal of Physics}\ }\textbf {\bibinfo {volume} {67}},\ \bibinfo {pages} {330} (\bibinfo {year} {1999})}\BibitemShut {NoStop}%
\bibitem [{\citenamefont {Ionicioiu}\ and\ \citenamefont {Terno}(2011)}]{terno2011}%
  \BibitemOpen
  \bibfield  {author} {\bibinfo {author} {\bibfnamefont {R.}~\bibnamefont {Ionicioiu}}\ and\ \bibinfo {author} {\bibfnamefont {D.~R.}\ \bibnamefont {Terno}},\ }\bibfield  {title} {\bibinfo {title} {Proposal for a quantum delayed-choice experiment},\ }\href {https://doi.org/10.1103/PhysRevLett.107.230406} {\bibfield  {journal} {\bibinfo  {journal} {Phys. Rev. Lett.}\ }\textbf {\bibinfo {volume} {107}},\ \bibinfo {pages} {230406} (\bibinfo {year} {2011})}\BibitemShut {NoStop}%
\bibitem [{\citenamefont {Masi}(2019)}]{masi}%
  \BibitemOpen
  \bibfield  {author} {\bibinfo {author} {\bibfnamefont {M.}~\bibnamefont {Masi}},\ }\href@noop {} {\bibinfo {title} {A review of modern which-way and delayed quantum erasing experiments: demystifying retro-causality and the point-particle myth.}} (\bibinfo {year} {2019})\BibitemShut {NoStop}%
\bibitem [{\citenamefont {Heisenberg}(1927)}]{heisenberg1927}%
  \BibitemOpen
  \bibfield  {author} {\bibinfo {author} {\bibfnamefont {W.}~\bibnamefont {Heisenberg}},\ }\bibfield  {title} {\bibinfo {title} {\"uber den anschaulichen inhalt der quantentheoretischen kinematik und mechanik},\ }\href {https://doi.org/10.1007/BF01397280} {\bibfield  {journal} {\bibinfo  {journal} {Zeitschrift f{\"u}r Physik}\ }\textbf {\bibinfo {volume} {43}},\ \bibinfo {pages} {172} (\bibinfo {year} {1927})}\BibitemShut {NoStop}%
\bibitem [{\citenamefont {von Weizs{\"a}cker}(1931)}]{weizsacker1931}%
  \BibitemOpen
  \bibfield  {author} {\bibinfo {author} {\bibfnamefont {C.~F.}\ \bibnamefont {von Weizs{\"a}cker}},\ }\bibfield  {title} {\bibinfo {title} {Ortsbestimmung eines elektrons durch ein mikroskop},\ }\href {https://doi.org/10.1007/BF01391035} {\bibfield  {journal} {\bibinfo  {journal} {Zeitschrift f{\"u}r Physik}\ }\textbf {\bibinfo {volume} {70}},\ \bibinfo {pages} {114} (\bibinfo {year} {1931})}\BibitemShut {NoStop}%
\bibitem [{\citenamefont {Einstein}\ \emph {et~al.}(1931)\citenamefont {Einstein}, \citenamefont {Tolman},\ and\ \citenamefont {Podolsky}}]{einstein1931}%
  \BibitemOpen
  \bibfield  {author} {\bibinfo {author} {\bibfnamefont {A.}~\bibnamefont {Einstein}}, \bibinfo {author} {\bibfnamefont {R.~C.}\ \bibnamefont {Tolman}},\ and\ \bibinfo {author} {\bibfnamefont {B.}~\bibnamefont {Podolsky}},\ }\bibfield  {title} {\bibinfo {title} {Knowledge of past and future in quantum mechanics},\ }\href {https://doi.org/10.1103/PhysRev.37.780} {\bibfield  {journal} {\bibinfo  {journal} {Phys. Rev.}\ }\textbf {\bibinfo {volume} {37}},\ \bibinfo {pages} {780} (\bibinfo {year} {1931})}\BibitemShut {NoStop}%
\bibitem [{\citenamefont {Hermann}(1935)}]{hermann1935}%
  \BibitemOpen
  \bibfield  {author} {\bibinfo {author} {\bibfnamefont {G.}~\bibnamefont {Hermann}},\ }\bibfield  {title} {\bibinfo {title} {Die naturphilosophischen grundlagen der quantenmechanik},\ }\href {https://doi.org/doi.org/10.1007/BF01491142} {\bibfield  {journal} {\bibinfo  {journal} {Abhandlungen der Fries'schen Schule}\ }\textbf {\bibinfo {volume} {6}},\ \bibinfo {pages} {75} (\bibinfo {year} {1935})},\ \bibinfo {note} {english translation in: "The Foundations of Quantum Mechanics in the Philosophy of Nature," Alkis Gounaris (Ed.), 2018.}\BibitemShut {Stop}%
\bibitem [{\citenamefont {Scully}\ and\ \citenamefont {Dr\"uhl}(1982)}]{scully1982}%
  \BibitemOpen
  \bibfield  {author} {\bibinfo {author} {\bibfnamefont {M.~O.}\ \bibnamefont {Scully}}\ and\ \bibinfo {author} {\bibfnamefont {K.}~\bibnamefont {Dr\"uhl}},\ }\bibfield  {title} {\bibinfo {title} {Quantum eraser: A proposed photon correlation experiment concerning observation and ``delayed choice" in quantum mechanics},\ }\href {https://doi.org/10.1103/PhysRevA.25.2208} {\bibfield  {journal} {\bibinfo  {journal} {Phys. Rev. A}\ }\textbf {\bibinfo {volume} {25}},\ \bibinfo {pages} {2208} (\bibinfo {year} {1982})}\BibitemShut {NoStop}%
\bibitem [{\citenamefont {Gaasbeek}(2010)}]{gaasbeek2010demystifying}%
  \BibitemOpen
  \bibfield  {author} {\bibinfo {author} {\bibfnamefont {B.}~\bibnamefont {Gaasbeek}},\ }\href {https://arxiv.org/abs/1007.3977} {\bibinfo {title} {Demystifying the delayed choice experiments}} (\bibinfo {year} {2010}),\ \Eprint {https://arxiv.org/abs/1007.3977} {arXiv:1007.3977 [cs.LG]} \BibitemShut {NoStop}%
\bibitem [{\citenamefont {Greenstein}\ and\ \citenamefont {Zajonc}(1997)}]{Greenstein1997}%
  \BibitemOpen
  \bibfield  {author} {\bibinfo {author} {\bibfnamefont {G.}~\bibnamefont {Greenstein}}\ and\ \bibinfo {author} {\bibfnamefont {A.~G.}\ \bibnamefont {Zajonc}},\ }\bibinfo {title} {The quantum challenge: Modern research on the foundations of quantum mechanics}\ (\bibinfo  {publisher} {Jones and Bartlett},\ \bibinfo {address} {Boston},\ \bibinfo {year} {1997})\ Chap.~\bibinfo {chapter} {2}\BibitemShut {NoStop}%
\bibitem [{\citenamefont {Qureshi}(2025)}]{qureshi2025enigma}%
  \BibitemOpen
  \bibfield  {author} {\bibinfo {author} {\bibfnamefont {T.}~\bibnamefont {Qureshi}},\ }\bibfield  {title} {\bibinfo {title} {The enigma of delayed choice quantum eraser},\ }\href {https://doi.org/10.12743/quanta.93} {\bibfield  {journal} {\bibinfo  {journal} {Quanta}\ }\textbf {\bibinfo {volume} {14}},\ \bibinfo {pages} {66} (\bibinfo {year} {2025})}\BibitemShut {NoStop}%
\bibitem [{\citenamefont {Qureshi}\ and\ \citenamefont {Rahman}(2012)}]{Qureshi2012}%
  \BibitemOpen
  \bibfield  {author} {\bibinfo {author} {\bibfnamefont {T.}~\bibnamefont {Qureshi}}\ and\ \bibinfo {author} {\bibfnamefont {Z.}~\bibnamefont {Rahman}},\ }\bibfield  {title} {\bibinfo {title} {Quantum eraser using a modified stern-gerlach setup},\ }\href {https://doi.org/10.1143/PTP.127.71} {\bibfield  {journal} {\bibinfo  {journal} {Progress of Theoretical Physics}\ }\textbf {\bibinfo {volume} {127}},\ \bibinfo {pages} {71} (\bibinfo {year} {2012})},\ \Eprint {https://arxiv.org/abs/https://academic.oup.com/ptp/article-pdf/127/1/71/19572854/127-1-71.pdf} {https://academic.oup.com/ptp/article-pdf/127/1/71/19572854/127-1-71.pdf} \BibitemShut {NoStop}%
\bibitem [{\citenamefont {Walborn}\ \emph {et~al.}(2002{\natexlab{b}})\citenamefont {Walborn}, \citenamefont {Terra~Cunha}, \citenamefont {P\'adua},\ and\ \citenamefont {Monken}}]{scully2002}%
  \BibitemOpen
  \bibfield  {author} {\bibinfo {author} {\bibfnamefont {S.~P.}\ \bibnamefont {Walborn}}, \bibinfo {author} {\bibfnamefont {M.~O.}\ \bibnamefont {Terra~Cunha}}, \bibinfo {author} {\bibfnamefont {S.}~\bibnamefont {P\'adua}},\ and\ \bibinfo {author} {\bibfnamefont {C.~H.}\ \bibnamefont {Monken}},\ }\bibfield  {title} {\bibinfo {title} {Double-slit quantum eraser},\ }\href {https://doi.org/10.1103/PhysRevA.65.033818} {\bibfield  {journal} {\bibinfo  {journal} {Phys. Rev. A}\ }\textbf {\bibinfo {volume} {65}},\ \bibinfo {pages} {033818} (\bibinfo {year} {2002}{\natexlab{b}})}\BibitemShut {NoStop}%
\bibitem [{\citenamefont {Burnham}\ and\ \citenamefont {Weinberg}(1970)}]{burnham1970}%
  \BibitemOpen
  \bibfield  {author} {\bibinfo {author} {\bibfnamefont {D.~C.}\ \bibnamefont {Burnham}}\ and\ \bibinfo {author} {\bibfnamefont {D.~L.}\ \bibnamefont {Weinberg}},\ }\bibfield  {title} {\bibinfo {title} {Observation of simultaneity in parametric production of dollar-quantum pairs},\ }\href {https://doi.org/10.1103/PhysRevLett.25.84} {\bibfield  {journal} {\bibinfo  {journal} {Physical Review Letters}\ }\textbf {\bibinfo {volume} {25}},\ \bibinfo {pages} {84} (\bibinfo {year} {1970})}\BibitemShut {NoStop}%
\bibitem [{\citenamefont {Kwiat}\ \emph {et~al.}(1995)\citenamefont {Kwiat}, \citenamefont {Mattle}, \citenamefont {Weinfurter}, \citenamefont {Zeilinger}, \citenamefont {Sergienko},\ and\ \citenamefont {Shih}}]{kwiat1995}%
  \BibitemOpen
  \bibfield  {author} {\bibinfo {author} {\bibfnamefont {P.~G.}\ \bibnamefont {Kwiat}}, \bibinfo {author} {\bibfnamefont {K.}~\bibnamefont {Mattle}}, \bibinfo {author} {\bibfnamefont {H.}~\bibnamefont {Weinfurter}}, \bibinfo {author} {\bibfnamefont {A.}~\bibnamefont {Zeilinger}}, \bibinfo {author} {\bibfnamefont {A.~V.}\ \bibnamefont {Sergienko}},\ and\ \bibinfo {author} {\bibfnamefont {Y.}~\bibnamefont {Shih}},\ }\bibfield  {title} {\bibinfo {title} {New high-intensity source of polarization-entangled photon pairs},\ }\href {https://doi.org/10.1103/PhysRevLett.75.4337} {\bibfield  {journal} {\bibinfo  {journal} {Physical Review Letters}\ }\textbf {\bibinfo {volume} {75}},\ \bibinfo {pages} {4337} (\bibinfo {year} {1995})}\BibitemShut {NoStop}%
\bibitem [{\citenamefont {Nielsen}\ and\ \citenamefont {Chuang}(2012)}]{nielsen2000quantum}%
  \BibitemOpen
  \bibfield  {author} {\bibinfo {author} {\bibfnamefont {M.~A.}\ \bibnamefont {Nielsen}}\ and\ \bibinfo {author} {\bibfnamefont {I.~L.}\ \bibnamefont {Chuang}},\ }\href {https://doi.org/10.1017/cbo9780511976667} {\emph {\bibinfo {title} {{Quantum Computation and Quantum Information}}}}\ (\bibinfo  {publisher} {Cambridge University Press},\ \bibinfo {year} {2012})\BibitemShut {NoStop}%
\end{thebibliography}%

\appendix
\section{Quantum State Evolution in the Delayed-Choice Experiment}
\label{app:state_evolution}

In this appendix, we present the complete calculations for the quantum state evolution across the optical components of the setups shown in Fig.~\ref{fig:interference},~\ref{fig:whichway} and~\ref{fig:delayedchoice}.
We first define a variable beam splitter with an internal parameter $\theta \in [0, \pi]$. For two input modes $\ket{a}$ and $\ket{b}$ yielding output modes $\ket{\gamma_1}$ and $\ket{\gamma_0}$, the unitary transformation is given by:
\begin{align}
\ket{a} &\rightarrow \cos\left(\frac{\theta}{2}\right) \ket{\gamma_1} + i\sin\left(\frac{\theta}{2}\right) \ket{\gamma_0}, \\
\ket{b} &\rightarrow i\sin\left(\frac{\theta}{2}\right) \ket{\gamma_1} + \cos\left(\frac{\theta}{2}\right) \ket{\gamma_0}.
\end{align}
 In our configuration, $\theta=\pi/2$ is chosen for $\text{BS}_1$, $\text{BS}_2$ and $\text{BS}_5$, whereas $\theta=0, \pi, \pi/2$ is chosen for $\text{BS}_3$ and $\text{BS}_4$, to generate case I, case II and case III, respectively in extended MZI shown in fig.~\ref{fig:delayedchoice}.
 After passing through the $\text{BS}_1$ the initial sate is defined as:
\begin{align}
\ket{a} & = \frac{  \ket{\gamma_1} + i\ket{\gamma_0}}{\sqrt{2}}, \\
\ket{b} & = \frac{  i\ket{\gamma_1} + \ket{\gamma_0}}{\sqrt{2}}
\end{align}
Entangled photon pair sources (EPPS) situated along arms $\gamma_0$ and $\gamma_1$ generate an entangled two-photon state (comprising idler modes $\ket{i}$ and signal modes $\ket{s}$)
\begin{equation}
\ket{\psi_0} = \frac{1}{\sqrt{2}} \left( \ket{i_0}\ket{s_0} + \ket{i_1}\ket{s_1} \right)\ket{D},
\end{equation}
where $\ket{D}$ is the initial state of the detectors.
The signal paths ($s_0, s_1$) are routed toward the beam splitter $\text{BS}_2$, whose output ports lead directly to detectors $D_1$ (mode $d_1$) and $D_2$ (mode $d_2$). The transformed  state in terms of mode $d_2$ and $d_1$ becomes:
\begin{equation}
\begin{aligned}
\ket{\psi_1} = \frac{1}{2} \Big[ &\ket{i_0} \left( -\ket{d_1} + i\ket{d_2} \right) \\
&+ \left( i\ket{d_1} - \ket{d_2} \right) \ket{i_1} \Big] \ket{D}.
\end{aligned}
\end{equation}
 Idler paths $i_0$ and $i_1$ are routed toward mirrors $M_1$ and $M_2$ which after reflection encounter the beam splitters $\text{BS}_3$ and $\text{BS}_4$, respectively (parameterized by $\theta$ as mentioned above). The state becomes: 
\begin{equation}
\begin{aligned}
\ket{\psi_2} = \frac{1}{2} \Bigg\{ &-\sin\left(\frac{\theta}{2}\right) \Big[ \ket{d_1}(i\ket{d_3} + \ket{d_4}) \\
&\quad + \ket{d_2}(\ket{d_3} + i\ket{d_4}) \Big] \\
&+ \cos\left(\frac{\theta}{2}\right) \Big[ \ket{d_1}(-\ket{i_0'} + i\ket{i_1'}) \\
&\quad + \ket{d_2}(i\ket{i_0'} - \ket{i_1'}) \Big] \Bigg\} \ket{D}.
\end{aligned}
\end{equation}

Finally, the modes $\ket{i_0'}$ and $\ket{i_1'}$ undergo recombination at $\text{BS}_5$, leading to the final state:
\begin{equation}
\begin{aligned}
\ket{\psi_{\text{final}}} = \frac{1}{2} \Bigg\{ 
&\ket{d_1} \left[ -\sin\left(\frac{\theta}{2}\right)(i\ket{d_3} + \ket{d_4}) \right. \\
&\quad \left. + \sqrt{2}i\cos\left(\frac{\theta}{2}\right)\ket{d_5} \right] \\
+ &\ket{d_2} \left[ -\sin\left(\frac{\theta}{2}\right)(\ket{d_3} + i\ket{d_4}) \right. \\
&\quad \left. - \sqrt{2}i\cos\left(\frac{\theta}{2}\right)\ket{d_6} \right] \Bigg\} \ket{D}.
\end{aligned}
\end{equation}

\begin{table}[htbp]
\centering
\renewcommand{\arraystretch}{1.8}
\begin{tabular}{|c|c|c|c|c|c|}
\hline
\multirow{2}{*}{Case} & \multirow{2}{*}{Signal Detector}& \multicolumn{4}{c|}{Idler Detector} \\
\cline{3-6}
 & & $D_3$ & $D_4$ & $D_5$ & $D_6$ \\
\hline
\multirow{2}{*}{I} & $D_1$ & $0$ & $0$ & $1/2$ & $0$ \\
\cline{2-6}
 & $D_2$ & $0$ & $0$ & $0$ & $1/2$ \\
\hline
\multirow{2}{*}{II} & $D_1$ & $1/4$ & $1/4$ & $0$ & $0$ \\
\cline{2-6}
 & $D_2$ & $1/4$ & $1/4$ & $0$ & $0$ \\
\hline
\multirow{2}{*}{III} & $D_1$ & $1/8$ & $1/8$ & $1/4$ & $0$ \\
\cline{2-6}
 & $D_2$ & $1/8$ & $1/8$ & $0$ & $1/4$ \\
\hline
\end{tabular}
\caption{Joint detection probabilities for coincident clicks between the signal detectors ($D_1$, $D_2$) and the idler detectors ($D_3$, $D_4$, $D_5$, $D_6$) for cases I, II, and III.}
\label{probabilities}
\end{table}

\textbf{Special  Cases}

The behavior of the setup is governed by the parameter $\theta$, which controls the transmission and reflection of beam splitters $\text{BS}_3$
 and $\text{BS}_4$. By varying $\theta$, the interferometer can be continuously tuned between different operational regimes.
 
\textbf{{Case I:  Interference and complementary interference ($\theta = 0$)}}
When $\theta = 0$, beam splitters $\text{BS}_3$ and $\text{BS}_4$ transmit completely. No photons are directed to output ports $D_3$ and $D_4$. The setup simplifies to the case I interferometer scheme shown in Fig.~\ref{fig:interference}, yielding:
\begin{equation}
\ket{\psi_{\text{I}}} = \frac{i}{\sqrt{2}} \left( \ket{\overline{D_1}}\ket{\overline{D_5}} - \ket{\overline{D_2}}\ket{\overline{D_6}} \right).
\end{equation}

\textbf{{Case II: Which-way measurement ($\theta = \pi$)}}
When $\theta = \pi$, beam splitters $\text{BS}_3$ and $\text{BS}_4$ act as total reflectors (mirrors). No photons reach output detectors $D_5$ and $D_6$. The setup corresponds to the case II  shown in Fig.~\ref{fig:whichway}, resulting in:
\begin{equation}
\begin{aligned}
\ket{\psi_{\text{II}}} = -\frac{1}{2} \Big[ &\ket{\overline{D_3}}\left(i\ket{\overline{D_1}} + \ket{\overline{D_2}}\right) \\
+ &\ket{\overline{D_4}}\left(\ket{\overline{D_1}} + i\ket{\overline{D_2}}\right) \Big].
\end{aligned}
\end{equation}

\textbf{{Case III: Quantum delayed choice ($\theta = \pi/2$)}}
When $\theta = \pi/2$, $\text{BS}_3$ and $\text{BS}_4$ operate as $50:50$ beam splitters. The system corresponds to Fig.~\ref{fig:delayedchoice} with the final state given as:
\begin{equation}
\begin{aligned}
\ket{\psi_{\text{III}}} = &-\frac{1}{2\sqrt{2}} \Big[ \ket{\overline{D_3}} \left( i\ket{\overline{D_1}} + \ket{\overline{D_2}} \right) \\
&\qquad\quad + \ket{\overline{D_4}} \left( \ket{\overline{D_1}} + i\ket{\overline{D_2}} \right) \Big] \\
&+ \frac{i}{2} \left[ \ket{\overline{D_5}}\ket{\overline{D_1}} - \ket{\overline{D_6}}\ket{\overline{D_2}} \right].
\end{aligned}
\end{equation}

The joint probabilities of detection at detectors D1 and D2 with respect to detectors D3, D4, D5 and D6 are shown in table \ref{probabilities} for all three cases.

\end{document}